\documentclass{osa-article}

\journal{osac}

\articletype{Research Article}

\begin{document}

\title{Reduction of laser intensity noise over 1 MHz band for single atom trapping}

\author{Yu Wang,\authormark{1,2} Kenneth Wang,\authormark{3,1,4} Eliot F. Fenton,\authormark{3} Yen-Wei Lin,\authormark{1,3,4} ,  Kang-Kuen Ni\authormark{1,3,4}, and Jonathan D. Hood,\authormark{5,6}}

\address{\authormark{1}Department of Chemistry and Chemical Biology, Harvard University, Cambridge, Massachusetts 02138, USA\\
\authormark{2}School of Physics, Peking University, Beijing 100871, China\\
\authormark{3}Department of Physics, Harvard University, Cambridge, Massachusetts 02138, USA\\
\authormark{4}Harvard-MIT Center for Ultracold Atoms, Cambridge, Massachusetts 02138, USA\\
\authormark{5}Department of Chemistry, Purdue University, West Lafayette, Indiana, 47907, USA\\
\authormark{6}Department of Physics and Astronomy, Purdue University, West Lafayette, Indiana, 47907, USA\\}
\email{\authormark{*}hoodjd@purdue.edu} 

\begin{abstract}
We reduce the intensity noise of laser light by using an electro-optic modulator and acousto-optic modulator in series. The electro-optic modulator reduces noise at high frequency (10 kHz to 1 MHz), while the acousto-optic modulator sets the average power of the light and reduces noise at low frequency (up to 10 kHz). The light is then used to trap single sodium atoms in an optical tweezer, where the lifetime of the atoms is limited by parametric heating due to laser noise at twice the trapping frequency. With our noise eater, the noise is reduced by up to 15 dB at these frequencies and the lifetime is increased by an order of magnitude to around 6 seconds. Our technique is general and acts directly on the laser beam, expanding laser options for sensitive optical trapping applications.  
\end{abstract}

\section{Introduction}
Low noise lasers are required in a variety of scientific applications including gravitational wave detection \cite{aasi2015advanced}, optical communication systems\cite{ahmed2008effect}, quantum key distribution\cite{madsen2012continuous}, and atom trapping \cite{blatt2015low}. Early approaches\cite{harb1994suppression, Taccheo:00, kane1990intensity} for intensity noise reduction in solid-state lasers directly fed back on the current of the pump diodes, which was effective for suppressing large relaxation oscillation peaks up to 1 MHz. However, if used with commercial lasers, this method is only suitable for those that can provide a fast feedback electronic path to the pump diode current. 

Other approaches that act on the light directly, independently of its source, have been developed for various frequency ranges of interest. For example, an approach designed for a gravitational wave detector uses feedback on an acousto-optic modulator (AOM) for frequencies below 10 kHz and an optical cavity for mode cleanup and frequencies above 1 MHz \cite{kwee2012stabilized}. The AOM technique has been shown to be effective to frequencies of 100's of kHz, limited by the bandwidth of the AOM \cite{blatt2015low}. To achieve noise reduction at even higher frequencies, electro-optical modulators (EOM) and semiconductor optical amplifiers (SOA) have been used. Feedback to an EOM has allowed for wideband intensity stabilization up to 10 MHz \cite{ivanov2007wide}, while feedforward has enabled stabilization into the GHz range \cite{michael2015broadband}. However, in both these cases, there is limited control over the DC setpoint of the laser light. Another approach involves injecting the laser light into a semiconductor optical amplifier, running in the saturated regime, which has been shown to reduce the intensity noise up to 50 MHz \cite{zhao2016broad}. However, these optical amplifiers introduce a large amount of broadband noise, extending out tens of nanometers, due to amplified spontaneous emission, requiring the use of a bandpass filter.

Here, we present a scheme for broadband noise reduction up to 1 MHz with large dynamic range. We implement a noise eater that combines an EOM and an AOM to independently suppress high (10 kHz--1 MHz) and low (DC--10 kHz) frequency intensity noise in series. We demonstrate our noise eater in an optical trapping application. Optically trapped ultracold atoms and molecules have emerged as a highly controllable and tunable platform for applications in precision measurement \cite{parker2015first, kozyryev2017precision, bothwell2019jila, kondov2019molecular, takamoto2005optical}, quantum simulation \cite{browaeys2020many}, and quantum computation \cite{levine2019parallel,graham2019rydberg,ni2018dipolar}. These optically trapped atoms and molecules are sensitive to the characteristics of the laser light forming the trap. In particular, laser intensity noise at twice the trap frequency can cause motional decoherence and heating \cite{gehm1998dynamics}. Thus, in many of these applications, laser noise at twice the trap frequency needs to be suppressed. In optical tweezers, the typical trap frequencies are in the 10 kHz - 1 MHz region \cite{kaufman2012cooling,thompson2013coherence, tuchendler2008energy, liu2019molecular, anderegg2019optical}, and we achieve up to 15 dB noise attenuation in this region. As a demonstration, we improve our sodium atom lifetime in the optical tweezer from 0.56 s to 5.8s, making it utilizable for further experiments in coherent molecule production\cite{zhang2020forming}, manipulation \cite{park2017second} and quantum gates\cite{ni2018dipolar, jaksch2000fast}.

\section{Experimental setup}
%
\begin{figure}[ht]
\centering
\includegraphics[width=12cm]{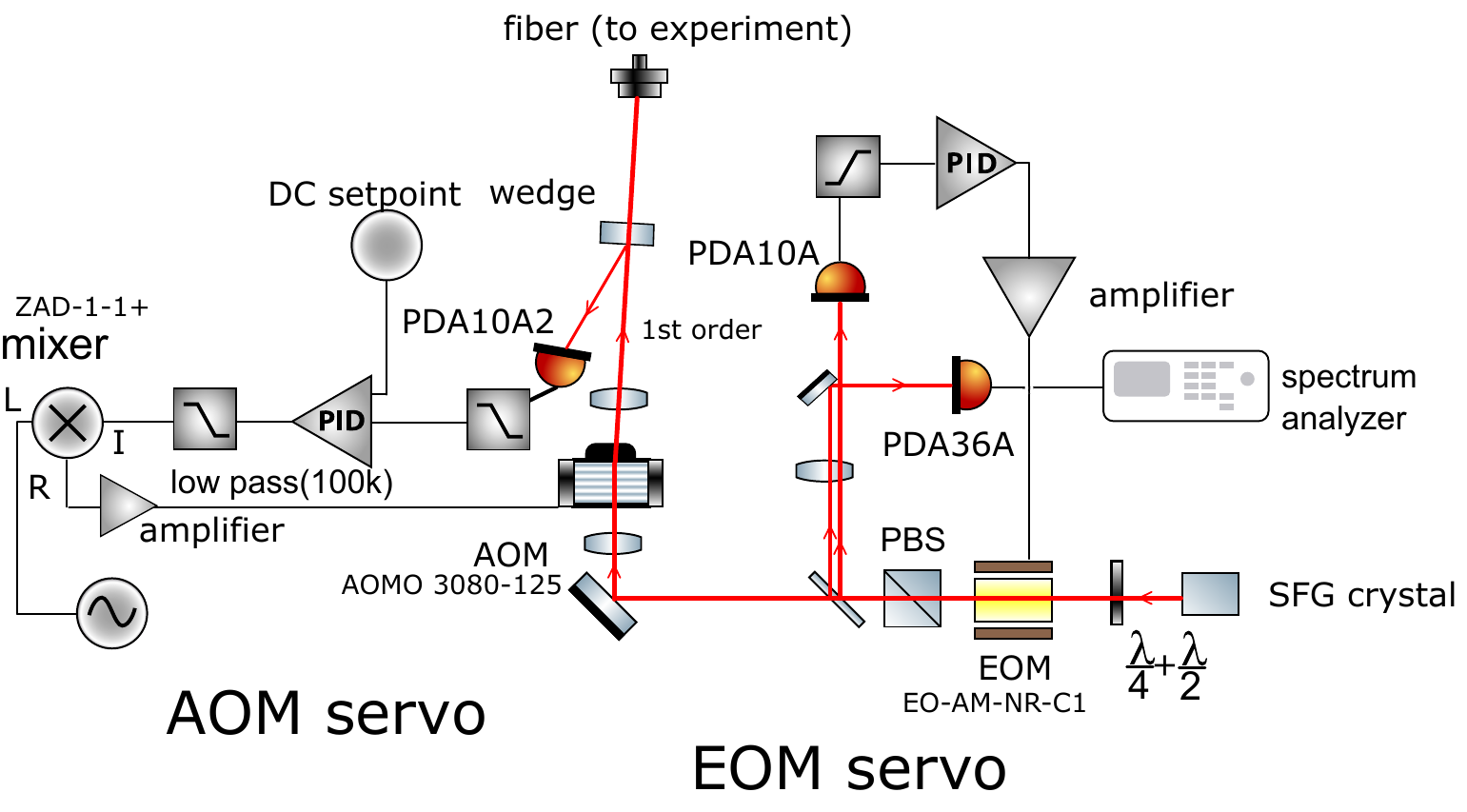}
\caption{Schematic of EOM and AOM servo. The 671 nm light (red line) exits the SFG crystal and is servo-ed by the EOM and the AOM successively. The AOM performs feedback on DC to 10kHz, and the EOM from 10kHz to 1MHz. }
\label{fig1}
\end{figure}

\begin{figure}[ht]
\centering
\includegraphics[width=10cm]{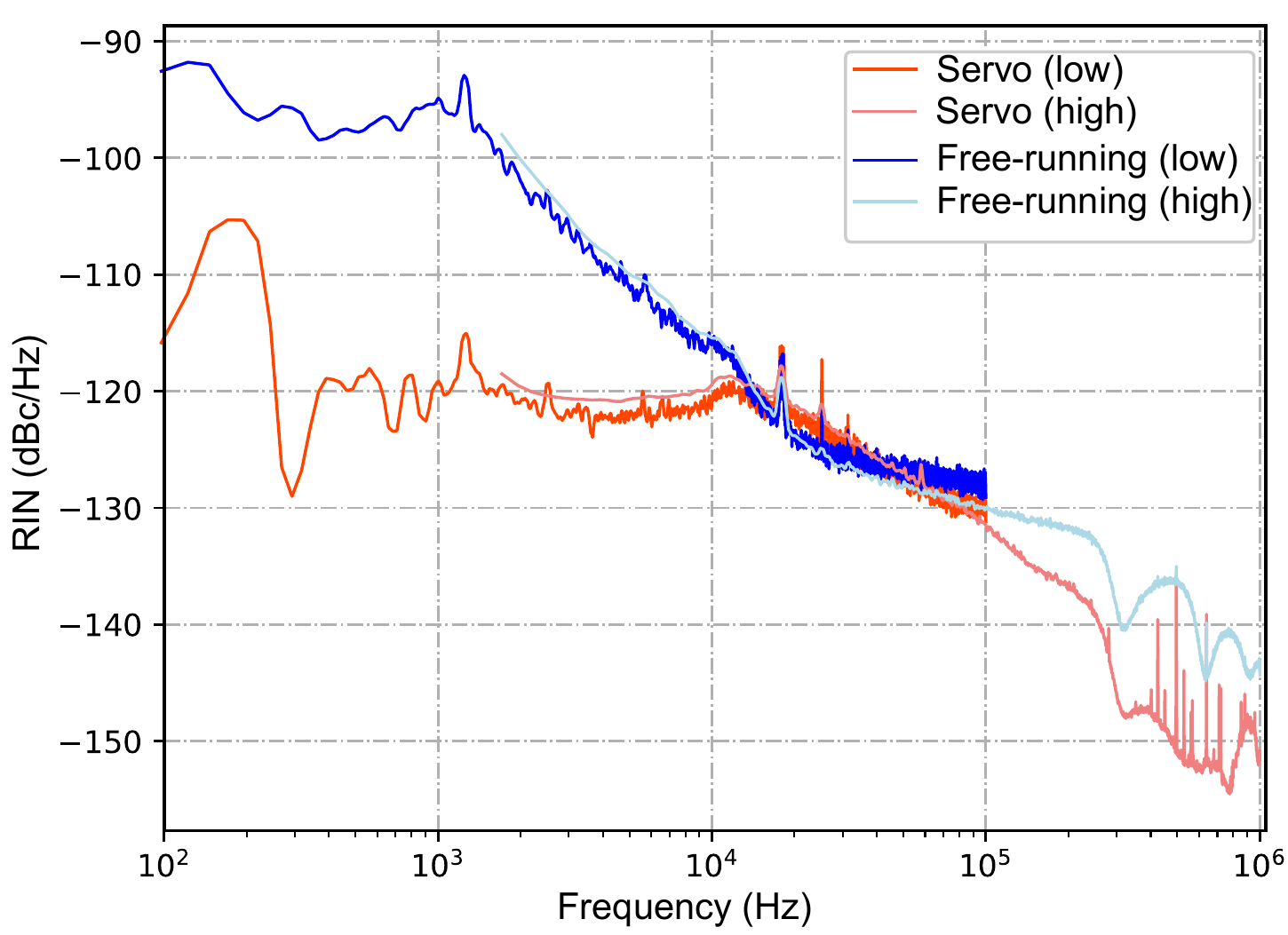}
\caption{The intensity noise spectrum. In the high-frequency region, the light blue and the light red curves shows the RIN with the noise eater off and on respectively. And in the low-frequency region, the blue and the red curves shows the RIN with the noise eater off and on respectively. In the low frequency region, the background, due to the dark current of the photodetector, is well below the measured signals. In the high frequency region, sharp peaks are due to noise in the AOM driver, and are consistent in both conditions. }
\label{fig:spectrum}
\end{figure}

In order to generate high power light at 671~nm for sodium atom trapping, we use sum frequency generation (SFG) of two commercially available fiber based lasers in the near-IR. We combine 1557~nm light from a NKT Photonics Erbium-doped fiber amplifier and 1178~nm light from a MPB Raman fiber amplifier in a MgO-doped periodically poled lithium niobate (MgO:PPLN) crystal from HC Photonics to get 671 nm light. Most of the noise in the output visible light comes from the 1178 nm laser, and is broadband at the -100 dBc$/$Hz level (see blue curve in Fig.~\ref{fig:spectrum}). The intensity noise at twice the atom's trapping frequency ($2f_{\text{axial}} = 76$~kHz, $2f_{\text{radial}}= 720 $~kHz) in the tweezer results in parametric heating, which shortens the trap lifetime.  

In order to reduce parametric heating and obtain shot to shot stability and control of our trap intensity, we require  feedback control across the DC-1 MHz range. A common approach is to modulate the RF power in an AOM, thus modulating the amount of power in a diffracted order, but feedback is limited to the few hundreds of kHz range due to the speed of sound in the AOM crystal. However, this problem can be alleviated by using an EOM in the amplitude modulator configuration, which can achieve a higher bandwidth.  In the amplitude modulator configuration, the polarization of light sent through the EOM crystal is rotated using a voltage, and the light is then attenuated by a polarizer. 
The difficulty in using EOMs comes from the large half-wave voltages, which is the voltage required to rotate the linear polarization by $90$ degrees. Typically, hundreds of volts are required  for full contrast. High voltage amplifiers with high gain are limited in bandwidth to several MHz, thus limiting servo performance.

By taking advantage of the strengths of each device independently, we can achieve significant noise reduction on the entire DC-1 MHz range. The AOM (Gooch and Housego AOMO 3080-125) handles low frequency well with easily achievable full dynamic range. The EOM (Thorlabs EO-AM-NR-C1) handles high frequency, where we expect the noise to be smaller, so large dynamic range and thus large voltages are unnecessary. 

In the experiment, we realize this concept with the setup shown in Fig.~\ref{fig1}. Feedback is implemented with a home built proportional-integral-derivative (PID) servo. To fully decouple the two servos from each other, we cascade them in series. The light first goes through the EOM, where high frequency noise is reduced, and then through the AOM, where its average power is set and low frequency noise is suppressed. Since our light is linearly polarized out of the nonlinear crystal, we use a quarter waveplate before the EOM and a polarizing beamsplitter (PBS) after the EOM to have linear servo response around zero input to the EOM. The maximum response is achieved with circularly polarized light as an input, but then transmission is sacrificed to 50$\%$. Empirically, we find that we have enough response to use elliptically polarized light such that the transmission is around 80$\%$. 
We then use a beam sampler to sample off two individual beams from the main beam and focus both onto photodetectors. We use one photodetector (Thorlabs PDA10A, bandwidth of 150 MHz for the feedback and the other one (PDA36A, bandwidth 10~MHz) for independent characterization of the servo performance. For the feedback photodetector signal, to ensure the EOM only acts on the high frequency components, we use a high-pass filter (a home built bias-T) with a 3dB cutoff of around 17~kHz to predominantly pass only the $ >17 $~kHz component to the servo. Without the need to address low frequency noise, we find that no integrator gain is required to get noise reduction. Some D gain is used to boost the gain at high frequencies. Furthermore, we pass the feedback signal through a  40~dB gain, 10~MHz bandwidth amplifier (Analog Devices AD8021) in order to increase the P gain beyond what our primary servo can deliver.  Importantly, the phase lags by less than 90 degree across the whole range, ensuring the system to not have positive feedback.

With the high frequency noise reduced, the light then passes through the AOM. We pick off some light from the first order diffracted beam and measure its intensity on a PDA10A2 photodetector (bandwidth of 150~MHz) and use it to feedback onto the RF power of the AOM. This is achieved with a separate home built PID servo, where the feedback signal is fed into the I port of a mixer to be mixed with the RF that drives the AOM. To ensure that the AOM servo does not introduce noise in the high frequency regime, we use a home built bias-T to filter out the high frequency in the photodetector signal as well as a 100 kHz low pass filter on the output of the servo. Here, we use a combination of P gain and I gain.  

\begin{figure}[ht]
\centering
\includegraphics[width=0.9\linewidth]{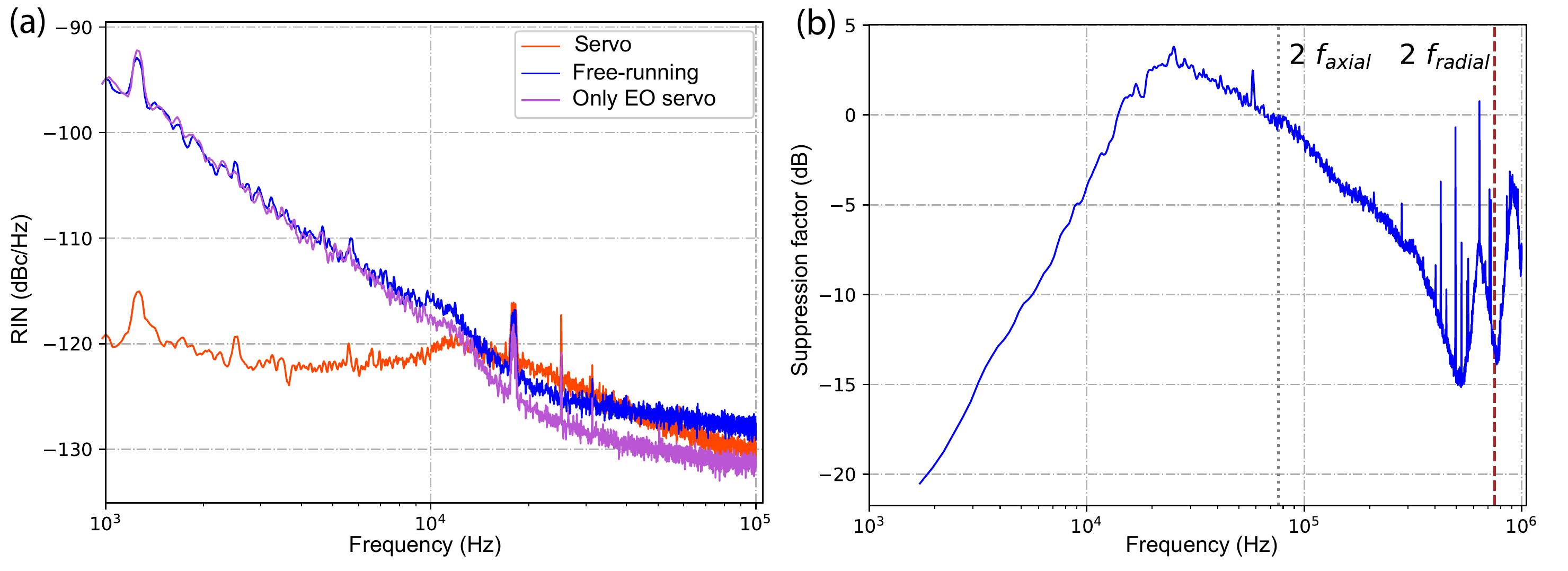}
\caption{(a) Intensity noise spectrum in the low-frequency region from 1 kHz--100 kHz. The blue curve shows the RIN of the free-running laser, without any servoing. The purple and red curves show the RIN with the AOM noise eater off and on respectively, while keeping the EOM noise eater on. (b) The noise reduction factor from 1 kHz to 1 MHz. Twice the axial and radial trapping frequency are shown as vertical dashed lines.  }
\label{fig3}
\end{figure}

\section{Results}
To characterize the performance of our broadband noise eater, we measure the light intensity on an independent photodetector, and use a low-noise SR560 preamplifier from Stanford Research Systems to improve the signal-to-noise ratio. We then measure the amplified signal and perform a Fourier transform with a Analog Discovery 2 from Digilent.

In Fig.~\ref{fig:spectrum}, we show how our noise eater performs. We see significant noise reduction, up to 15dB throughout the 50 kHz - 1MHz band, and up to 20 dB in the 100 Hz - 10 kHz band. We determine the performance of the AOM at low frequency by measuring the light on an independent photodetector after the AOM. In Fig.~\ref{fig3}(a), we keep the EOM servo active, while comparing the results with and without the AOM servo active. Turning on the AOM servo reduces the noise in the 20 Hz - 10 kHz range, which is unaddressed by the EOM servo. In the 10 kHz - 50 kHz range, by design, due to the bias T cutoff frequency of 17 kHz, both servos do not reduce the noise in this range, and there is at most a 1 dB increase in the noise. We note that both servos have no appreciable effect on the noise above 1 MHz. \\

We further characterize our servo performance by measuring the lifetime of a sodium atom in an optical tweezer formed from the 671 nm light, which is red detuned from the Na 3s to 3p transition at about 589 nm.  The light is focused to a diffraction limited spot with a 0.55 numerical aperture microscope objective. In this system, the lifetime of the atom is limited by parametric heating at twice the trap frequency in any of the three axes. With a lower noise laser source, we see an increase in the atom lifetime.

\begin{figure}[ht]
\centering
\includegraphics[width=0.9\linewidth]{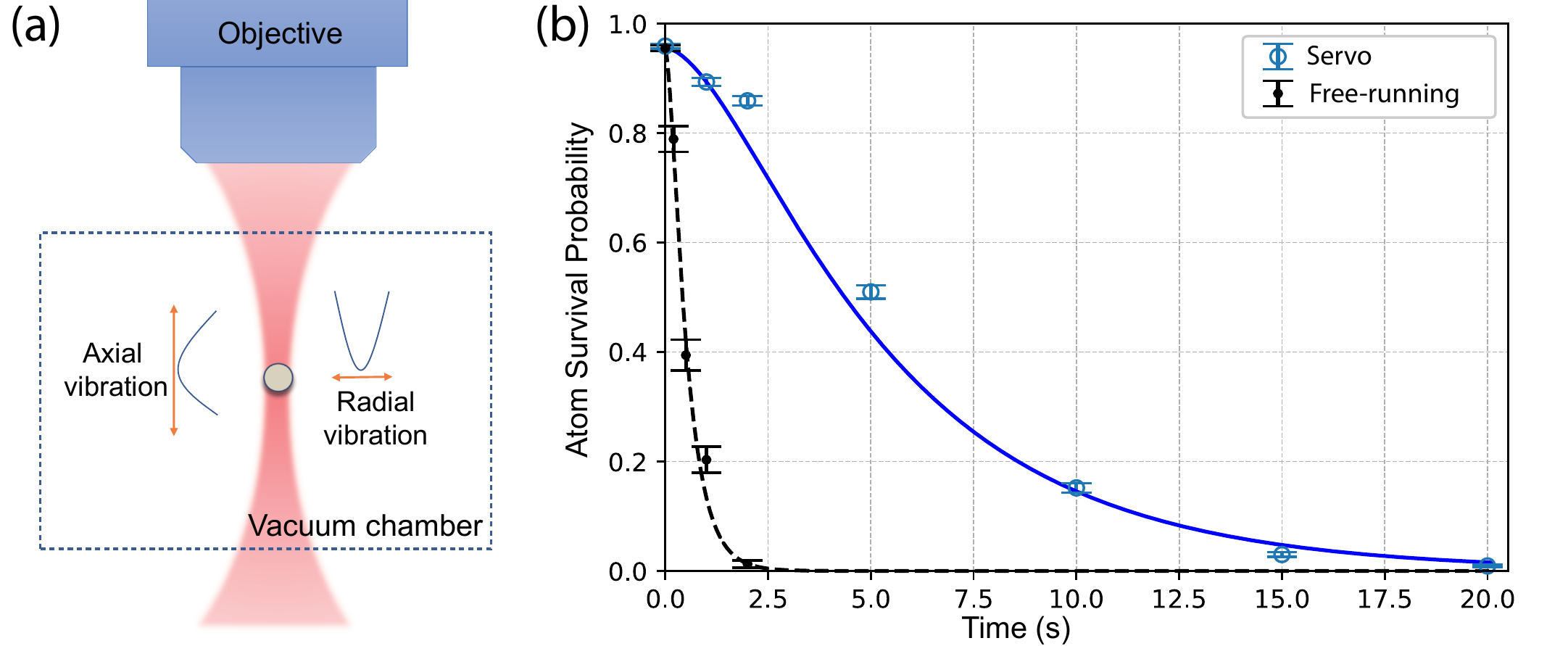}
  \caption{(a) Optical tweezer setup.The dotted rectangle shows the vacuum chamber. The red beam shows the Gaussian laser beam from the objective to trap the atom, which is shown with the white ball in the center. The trapped atom in the tweezer moves like a 3D harmonic oscillator with a low axial frequency and a high radial frequency. (b) The survival probability of the atom as a function of time. The blue and black points are the experimental data when the servo is active or not. The blue solid line and the black dashed line are the corresponding fits.  }
  \label{fig4}
 \end{figure}

We stochastically load single sodium atoms into an optical tweezer from a magneto-optical trap (MOT) \cite{hutzler2017eliminating}. We then hold the atom for a variable amount of time in the 671 nm optical tweezer, and measure the survival probability with and without the noise eater active. The results in Fig.~\ref{fig4}(b) show an increase in the atom lifetime when the noise eater is used.

For a more quantitative analysis of the data, we fit the survival probabilities following the model in Ref.~\cite{gehm1998dynamics}. We assume an initial Maxwell-Boltzmann energy distribution, instead of a Bose-Einstein distribution, since we are loading one atom at a time. We then evolve the energy distribution to time $ t $ using the Fokker-Planck equation with a heating rate, $ \Gamma $. Then, to obtain the survival probability of the atoms, we calculate the portion of the distribution that is lower in energy than the trap depth. From this procedure, we fit the heating rate, and the ratio of the initial temperature to the trap depth. We measure the trap power, beam waist, and convert to a trap depth \cite{arora2007magic}. Using this trap depth of 5.7 mK, we can then extract an initial temperature, $ T_0 $.

We fit both the servo active and free running cases shown in Fig.~\ref{fig4} together, extracting a common $T_0$ along with heating rates, $\Gamma_{\text{servo}}$ and $ \Gamma_{\text{free}} $:
\begin{equation} \label{data_heating}
    \begin{split}
        &T_0 = 0.50\pm0.02 \text{ mK}\\
        &\Gamma_{\text{free}} = 3.50\pm 0.03 \text{ s}^{-1}\\
        &\Gamma_{\text{servo}} = 0.348\pm 0.003 \text{ s}^{-1} \\
    \end{split}
\end{equation}
The noise eater reduces the heating rate by a factor of 10, which increases the $1/e$ lifetime from 0.56s to 5.8s. 

As a further check to the validity of the model, we can also extract the heating rates directly from the intensity spectra. The relationship between heating rate along axis $ i $ and intensity noise is given in Ref.~ \cite{gehm1998dynamics} by 
\begin{equation}
\Gamma_i=\pi^2\times f_{i}^2\times S_k(2f_{i})
\label{eq:gamma}
\end{equation}
where $S_k(2f_{i})$ is the one-sided power spectrum of the fractional fluctuation in the light intensity of the tweezer at twice the trap frequency of axis $ i $. Assuming ergodicity, the heating rate in the 3D trap is then given by the arithmetic average of the heating rate in each direction, $\Gamma=(\Gamma_x+\Gamma_y+\Gamma_z)/3$  \cite{gehm1998dynamics}. To extract these heating rates, we measure our trap frequencies using Raman sideband spectroscopy\cite{PhysRevA.97.063423, PhysRevLett.75.4011}. The trap frequencies in our cylindrical trap are  $
f_{\text{radial}}=f_x=f_y= 360\pm 3$ kHz, $f_{\text{axial}} =f_z= 38\pm 1 $kHz.
Using the noise spectra, we extract the expected heating rates for the servo active and free running conditions $\Gamma_{\text{free}} = 5.1\pm0.9 \text{ s}^{-1}$ and $\Gamma_{\text{servo}} = 0.3\pm 0.1 \text{ s}^{-1}$.
%
%
These values are in reasonable agreement with the heating rates we extracted from the survival data (Eq.~\ref{data_heating}), indicating that parametric heating from the tweezer light is the dominant mechanism limiting the atom lifetime. A separate independent measurement in a different wavelength optical tweezer also indicated that the lifetime is not limited by collisions with the background gas. 

\section{Conclusion}
In conclusion, we developed a scheme for broadband intensity noise reduction in lasers. In our case, we are able to reduce the intensity noise of an optical tweezer up to 15 dB over a 1~MHz band with the noise eater we designed. The EOM and AOM feedback loops work separately on the intensity noise in the high frequency region and low frequency region. We have demonstrated stable trapping of a single sodium atom in this 671 nm optical tweezer. With the noise eater, we achieve about a 10 times lower heating rate, leading to a lifetime longer than 5 s. The heating rates extracted from the lifetime data are in good agreement with the results calculated from the noise spectra based on the parametric heating mechanism. 
With this servo scheme, we show that atom loss in optical tweezers due to parametric heating is reduced. This reduction in atom loss is critical for future experimentation in quantum computation and simulation with neutral atom tweezer arrays, where atom loss becomes exponentially worse with system size. 

\section{Acknowledgement}
We would like to thank Yichao Yu and Ming-Guang Hu for helpful discussions. This work is supported by ARO
DURIP (W911NF1810194), the NSF-CUA (PHY-1734011) and the Camille and Henry
Dreyfus Foundation (TC-18-003)  K. W. is supported by an NSF GRFP fellowship. 

\bibliography{ref}

\end{document}